\documentclass[12pt, a4paper]{article}
\usepackage{hyphenat}
\usepackage[onelanguage,linesnumbered,ruled]{algorithm2e}
\usepackage{bbm}
\usepackage{graphicx}%
\usepackage{multirow}%
\usepackage{titling}
\usepackage{hyperref}
\usepackage[margin=.75in]{geometry}
\usepackage[numbers]{natbib}
\usepackage{amsmath,amssymb,amsfonts}%
\usepackage{amsthm}%
\usepackage{mathrsfs}%
\usepackage[title]{appendix}%
\usepackage{xcolor}%
\usepackage{textcomp}%
\usepackage{manyfoot}%
\usepackage{booktabs}%
\usepackage{algorithmicx}%
\usepackage{algpseudocode}%
\usepackage{listings}%
\usepackage{babel}%
\usepackage{subfig}%
\usepackage{lmodern}

\title{Siamese networks for Poincar\'e embeddings and the reconstruction of evolutionary trees}

\usepackage{authblk}  

\author[1]{Ciro Carvallo}
\author[2,3]{Hernán Bocaccio}
\author[2,3]{Gabriel B. Mindlin}
\author[1,4]{Pablo Groisman}

\affil[1]{Departamento de Matemática, Facultad de Ciencias Exactas y Naturales, Universidad de Buenos Aires}
\affil[2]{Departamento de F\'isica, Facultad de Ciencias Exactas y Naturales, Universidad de Buenos Aires}
\affil[3]{CONICET - Universidad de Buenos Aires, Instituto de Física Interdisciplinaria y Aplicada (INFINA)}
\affil[4]{CONICET - Universidad de Buenos Aires, Instituto de Matem\'atica Luis A. Santal\'o (IMAS)}

\affil[*]{Email: cirocarvallo@gmail.com (C. Carvallo), hbocaccio@gmail.com (H. Bocaccio), gabo@df.uba.ar (G.B. Mindlin),  pgroisma@dm.uba.ar (P. Groisman)}

\date{}

\begin{document}
\maketitle
\abstract{We present a method for reconstructing evolutionary trees from high-dimensional data, with a specific application to bird song spectrograms. We address the challenge of inferring phylogenetic relationships from phenotypic traits, like vocalizations, without predefined acoustic properties. Our approach combines two main components: Poincaré embeddings for dimensionality reduction and distance computation, and the neighbor joining algorithm for tree reconstruction. Unlike previous work, we employ Siamese networks to learn embeddings from only leaf node samples of the latent tree. We demonstrate our method's effectiveness on both synthetic data and spectrograms from six species of finches.}

\section{Introduction}
\label{intro}

Animal classification systems are based on evolutionary relationships between different organisms, known as phylogeny. This approach allows us to organize species in a way that reflects our understanding of how they evolved from common ancestors. Phylogenetic trees are diagrams that graphically represent these evolutionary relationships between organisms. In these representations, the species of interest are placed at the tips of branches that emerge from a point representing a common ancestor. The natural mathematical object associated to this situation is a tree (a graph with no cycles) with a root (the common ancestor). It is important to note that the hypotheses regarding how different species may have descended from a common ancestor are typically based on physical traits (which are therefore interpretable) or directly on DNA sequences.

In this context, one can attempt to infer phylogenetic distances based on phenotypic traits. For instance, in the case of birds, one might consider the possibility of inferring these distances based on their songs \cite{CATE2004296, Mason2016, Chen2020, Martens1996, Wimberger1996, Price2002, Beecher2005, Tachibana2014}. If that is the case, what set of acoustic properties would allow us to predict phylogenetic distances? In this study, we propose a method based on dimensionality reduction (embeddings) and tree reconstruction to process sonograms (i.e., graphical representations of bird songs) in order to test the hypothesis that it is possible to predict phylogenetic distances from bird songs without predefining the relevant properties.

On one hand, we address the general problem of reconstructing latent trees from images, and then we apply this method to a recently discussed example in biology, where pre-defined acoustic properties were used in conjunction with AI techniques to infer phylogenetic distances among a set of songbird species \cite{moreira2023hyperbolicvseuclideanembeddings} \cite{rivera_woolley2023}.

Our approach consists of two main components:

\begin{enumerate}
 \item Finding an embedding of the sonograms in a space where we can compute meaningful distances.
 \item Reconstructing the tree based on these distances.
\end{enumerate}

Each of the components is based on well-known methods: Poincaré embeddings \cite{poincareembeddings} to obtain the desired distances and the neighbor joining algorithm \cite{Saitou1987TheNM, bioinformatics} to reconstruct the tree. Poincaré embeddings, which map data into hyperbolic space, have shown promise in representing hierarchical structures and obtaining simultaneously distances in the embedded spaces that reflect similarities. Hyperbolic geometry allows for the efficient representation of tree-like structures (see \cite{Krioukov2010HyperbolicNetworks, Boguna2009NavigabilityNetworks, Mikolov2013DistributedCompositionality, Liu2019HyperbolicNetworks, Sala2018RepresentationEmbeddings, Chami2019HyperbolicNetworks, Atigh2021HyperbolicPrototypes, Liu2020HyperbolicRecognition, Khrulkov2019HyperbolicEmbeddings, Guo2022ClippedClassifiers, GhadimiAtigh2022HyperbolicSegmentation, Mathieu2019ContinuousAuto-encoders, Nagano2019ALearning, Skopek2020Mixed-curvatureAutoencoders, Lin2022ContrastiveClustering, Yue2023HyperbolicLearning, Ge2022HyperbolicObjects, moreira2023hyperbolicvseuclideanembeddings} and in particular the introduction in \cite{moreira2023hyperbolicvseuclideanembeddings} for an account of the state of the art), making it particularly suitable for our task. The neighbor joining algorithm, a well-established method in phylogenetics, is then applied to the embedded data to reconstruct the tree.

Unlike previous work in hierarchical reconstruction from embeddings \cite{poincareembeddings}, our method deals with the challenge of having only samples of leaves from the latent tree. To address this, we employ Siamese networks, which are particularly well-suited for learning similarity metrics between pairs of samples. This approach differs from the loss function considered in \cite{poincareembeddings}, which assumes access to nodes at all the levels of the latent hierarchy during training. Siamese networks have been previously used to embed sonograms in Euclidean space \cite{BISTEL2022112803}.

We demonstrate the effectiveness of our approach on both synthetic data with known hierarchical structure and real-world bird song spectrograms. For the latter, we focus on six species of estrildid finches: {\em Stizoptera bichenovii}, {\em Poephila acuticauda}, {\em Lagonosticta senegala}, {\em Lonchura striata}, {\em Uraeginthus bengalus}, and {\em Amandava subflava}. These species were chosen based on a recent study comparing their phylogenetic organization with one emerging from pre-selected acoustic properties of their songs \cite{rivera_woolley2023}. Our method offers several advantages over traditional approaches:

\begin{enumerate}
 \item It does not require pre-specification of acoustic or other type of properties, instead learning relevant features directly from the spectrograms.
 \item It provides a general framework for reconstructing trees from high-dimensional representations of their leaves, applicable beyond bird song analysis.
 \item The use of hyperbolic geometry allows for more efficient representation of hierarchical structures compared to Euclidean embeddings.
\end{enumerate}

The rest of this paper is organized as follows: Section \ref{poincare} provides background on hyperbolic geometry and Poincaré embeddings. Section \ref{Siamese} describes our Siamese network architecture for learning embeddings. Section \ref{tree} details with the neighbor joining algorithm and our tree reconstruction process. Section \ref{results} presents our results on both synthetic and bird song data, comparing our reconstructed trees to known phylogenetic relationships. Finally, Section \ref{conclusion} discusses the implications of our findings and potential directions for future research.

This work contributes to the growing body of research at the intersection of artificial intelligence, machine learning and evolutionary biology, offering a novel method for inferring evolutionary relationships from complex, high-dimensional data such as spectrograms. By bridging the gap between behavioral traits and phylogenetic reconstruction, our approach has the potential to provide new insights into the evolution of animal communication and behavior. 

\section{Poincar\'e Embeddings}
\label{poincare}
\subsection{Hyperbolic Geometry and the Poincar\'e ball model}

Hyperbolic geometry is a non-Euclidean geometry in which the parallel postulate is replaced with:

\medskip

\begin{quotation}
{\em For any given line $R$ and point $P$ not on $R$, in the plane containing both line $R$ and point $P$ there are at least two distinct lines through $P$ that do not intersect $R$.}
\end{quotation}

\medskip

Hyperbolic spaces have begun to receive attention in the machine learning community as they are suitable for modeling data with hierarchical latent structure (see \cite{poincareembeddings} and references therein). Heuristically, they can be thought as continuous version of trees.

The Poincar\'e disk (ball) model uses the interior of a Euclidean unit disk (ball) as the representation space, $\mathcal D = \{\left(x,y\right) \colon x^2+y^2 < 1\}$. The boundary $\partial \mathcal D = \{(x,y)\colon x^2 + y^2 = 1\}$ of $\mathcal D$ does not belong to the disk. The distance between points $u, v \in \mathcal D$ is defined by
\begin{equation}
\label{eqn:poincare distance}
 d\left(u, v\right) = \textrm{arcosh} \left( 1 + \frac{2 \|u - v\|^2}{(1 -
 \|u\|^2)(1 - \|v\|^2)}\right).
\end{equation}
The $d-$dimensional Poincar\'e ball model is obtained similarly by replacing the two dimensional disk with a $d-$dimensional ball. The Poincar\'e ball is a metric space and moreover, it can be equipped with a Riemannian metric tensor to obtain a Riemannian manifold, but we will not require this.

When $u$ is the center of the circle and $v$ is at Euclidean distance $r$ from the origin we get
\begin{equation}
d(0,v) =  \textrm{arcosh} \left({\frac {1+r^2}{1-r^2}}\right) = \ln \left({\frac {(1+r)^2}{1-r^2}}\right)=2 \textrm{artanh}\left( r\right).
 \label{eq: distancia r}
\end{equation}
Note that the distance grows unbounded as $v$ approaches to the boundary and, consequently, the hyperbolic distance from a point in the interior of $\partial \mathcal D$ to the boundary of the disk is infinite.

\subsection{Poincar\'e Embeddings}

Adequate embeddings (representations) of symbolic data in different type of spaces and by different means have become a key point for many learning tasks. The most prominent examples are word embedding and their use to train Large Language Models \cite{mikolov2013efficient, pennington2014glove, bojanowski2017enriching, peters2018deep, devlin2018bert, radford2018improving, liu2019roberta, vaswani2017attention}.

In \cite{poincareembeddings} the authors suggest to use the Poincar\'e ball as a suitable space to look for a good embedding for data with a latent hierarchical structure.

\begin{figure}[ht]
\centering
\includegraphics[scale=0.5]{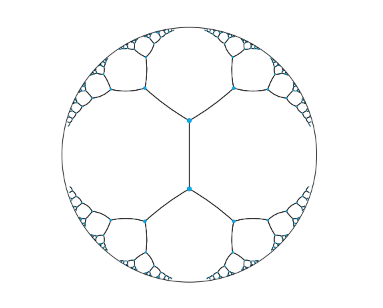}
\caption{Embedding of a tree with branching factor 2 into the Poincar\'e disk. Image taken from \cite{poincareembeddings}.}
\label{fig: hierarchichal embedding}
\end{figure}

If we consider data with a hierarchy like a tree, then the Euclidean space is not ideal. For example, a tree with a branching factor $b$ has $(b+1)b^{\ell-1}$ nodes at level $\ell$ and $\frac{(b + 1)b^\ell -2}{b-1}$ nodes at levels less than or equal to $\ell$. Thus, as the levels of the tree grow, the number of nodes grows exponentially. If we want to embedded such a tree in Euclidean space in such a way that the distances between nodes is adequately represented by Euclidean distances, we need the dimension of the space to be as large as the number of levels in the tree minus one.

Remarkably, we can represent our tree within two dimensional hyperbolic space by means of the following algorithm: we place the root of the tree at the center and their children equidistantly. Then, we repeat this procedure recursively in each branch. In figure \ref{fig: hierarchichal embedding}, we can visualize this embedding. In it, all points are equidistant from their parents. This type of construction is possible because in hyperbolic space we can accommodate as many equidistant points as desired by settling them close enough to the boundary.

Thus, if the root of a tree is placed at the origin, the relative distance to other nodes is small. However, the leaves can be located near the boundary to guarantee that the hyperbolic distance coincides with the graph distance in the tree.

In view of this, we are going to look for an embedding of sonograms from bird songs into the Poincar\'e disk (ball) with the hope that such embeddings can give information about a latent hierarchical structure determined by phylogenetics. Previously, to address the performance of the method, we will deal with synthetic data composed of images with a latent hierarchical structure.

The goal of computing embeddings for images coming from both synthetic data and spectrograms, is to obtain a representation in a space in which distances between points are meaningful. They simplify the representation of images by reducing dimensionality drastically while preserving relationships between them.

The Poincar\'e ball is used as the embedding space for both our synthetic data and spectrograms data because they both have the desired hierarchical structure and are able accommodate easily such hierarchy.

In the synthetic data, they represent the leaves of a tree constructed by branching areas of a square into zeros or ones. Spectrograms are the songs of different species that can be related through a phylogenetic tree; remember that the hypothesis is that spectrograms are the leaves of the evolutionary tree.

We remark that our inspiration to use Poincaré embeddings comes from \cite{poincareembeddings} but our situation is different to that as well us our algorithm to find the embedding. On the one hand, in \cite{poincareembeddings} the algorithm is based on information about hyponomy relations between pairs of symbols and this is the only information that is used. On the other hand, in \cite{poincareembeddings} it is implicitly assumed that the data fulfills the Poincaré disk (data at all levels of hierarchy). In our case, given a pair of sonograms, besides the information about if they are coming from the same specie or not, we have a (high dimensional and poor) notion of similarity between the sonograms their self. On the other hand, our data corresponds only to leaves of the tree, so that we don't expect to fulfill the whole disk but to be close to the boundary.

For these reasons, instead of following the approach of \cite{poincareembeddings} we are going to proceed to find the embeddings by means of Siamese networks and a different loss function. That is the purpose of the next section.

\section{Siamese Networks}
\label{Siamese}

A Siamese network is a neural network composed of two or more identical sub-networks that share the same weights and architecture, joined by a separate output layer. They are used to find the similarity between two comparable objects. Among other tasks, they have been used successfully for face recognition and signature verification \cite{Chicco2020, Bromley1994, Chopra2005, Taigman2014}.

The architecture of a Siamese network typically consists of a convolutional backbone followed by a dense layer. The input images are passed through the backbone and the output features are passed to a dense layer, which calculates a vector of dimension $n$. This vector is then passed to the output layer, which calculates the Euclidean distance between the feature vectors of the input images.

A typical loss function used for training a Siamese network is the contrastive loss function,
\begin{equation}
\label{loss}
    L(w) = \sum_{(i,j) \in \mathcal T}\frac{(1 - y_{ij}) }{2} d_{ij}^{2} + \frac{y_{ij}}{2} \left[ \text{max}(0, m - d_{ij}) \right]^{2}
\end{equation}
where $y_{ij}$ is the label indicating whether the images $x_i$ and $x_j$ are similar ($y_{ij} = 0$) or dissimilar ($y_{ij} = 1$), and $d_{ij}$ is the distance between the feature vectors of the input images, which is usually and $L^p-$norm in Euclidean space. The training set $\mathcal T$ is a subset of the product space of all possible pairs that is suitable chosen. The variable $w$ stands for the weights of the network.

The term $\frac{(1 - y) }{2} d^{2}$ penalizes similar images that have a large distance between them, and $\frac{y}{2} \left[ \text{max}(0, m - d) \right]^{2}$ penalizes dissimilar images whose distance is less than a margin $m$.

Siamese networks are trained with pairs of similar and dissimilar images that constitute the set $\mathcal T$. Their output can be used to measure the similarity between two images.

Since we are looking for an embedding in hyperbolic space, in this work we propose to replace the usual choice of $d_{ij}$ with the distance in the Poincar\'e ball.

\section{Tree Reconstruction}
\label{tree}
In this section we review the neighbor joining algorithm for tree reconstruction. The input of the algorithm is a set of (compatible) distances between leaves of the (latent) tree. The output is a tree in which the distance between leaves are the ones given by the input.

\subsection{Phylogenetic Tree}

Phylogenetic trees allow visualizing the evolutionary development of different species, subspecies, or populations called \emph{taxa} and the relationships between them. Each of the leaves of these trees represents different taxa, the interior vertices represent common ancestors from which they derive. If there is a common ancestor to all taxa, it is located at the root of the tree.

The most recent taxa are represented as leaves of the trees. Given a leaf $j$, the only node to which it is connected is called the parent of $j$. Also, the edge connecting a leaf to its parent is called a branch.

In the reconstruction of phylogenetic trees, the aim is typically to relate taxa that are currently present. Information about these is available, but not about those represented by internal nodes. The goal is to reconstruct evolutionary relationships from the distances between the terminal taxa. Distances information can be obtained in various ways. In the case of the bird species analyzed in this work, distance information arises from differences and similarities between their songs and is measured through the embedding of the sonograms in the Poincar\'e ball.

\subsection{Distance Matrices}

A matrix $D \in \mathbb R^{n \times n}$ with nonnegative entries is used to represent the distances between different taxa. For such a matrix $D$ we will say that it is a {\em distance matrix} if it is symmetric,  and satisfies the triangular inequality, $D_{i,k} \leq D_{i,j} + D_{j,k}$ for all $1\le i,j,k \le n$.

To represent the evolutionary distance between taxa, a length is assigned to each edge. The length of a path within the tree is defined as the sum of the distances of each edge. Thus, for a tree $T$, the evolutionary distance between two present taxa $i$ and $j$, called $d_{i,j}(T)$, corresponds to the length of the shortest path from $i$ to $j$.

\subsection{Neighbor joining Algorithm}

One of the most commonly used methods for reconstructing evolutionary trees is the neighbor joining (NJ) algorithm proposed by Saitou and Masatoshi \cite{Saitou1987TheNM}. We briefly review the algorithm following \cite{bioinformatics}. Given a distance matrix $D$ that represents distances between leaves of an unknown tree, the algorithm selects pairs of leaves that are inferred to be neighbors and substitutes them with a single leaf, thus reconstructing a tree by repeating the process recursively. A natural way to do this is to choose a pair of leaves with minimum distance and to assume that they have a common parent. However finding a minimum in the distance matrix does not guarantee finding a pair of siblings in the tree. Hopefully, it is possible to transform this matrix into another one whose minimum element is indeed a pair of siblings. Proceeding recursively we can reconstruct the whole tree.

The NJ algorithm produces an unrooted phylogenetic tree. However, knowing the root is of great importance for interpreting evolutionary relationships, as it represents the common ancestor to all taxa of the tree and provides the tree with an evolutionary path of the species.

There are several methods to place the root of a tree, the most used one being \textit{outgroup rooting}. In this case we need external (biological) knowledge to identify subgroups in the tree. Since we want to have an algorithm which is free of external knowledge, we opt to use \textit{midpoint rooting} \cite{midpoint}, which is only based on the distances calculated by the tree reconstruction algorithm.

\section{Results}
\label{results}
In this section we show how to apply our method. First, we will work with synthetic data, generated with a controlled ``evolutionary'' algorithm. Then we will apply our method to sonograms computed from a public data base (Xeno-canto). Our method is implemented with the following algorithm:

\begin{algorithm}
    \SetKwInOut{Input}{Input}
    \SetKwInOut{Output}{Output}
    \Input{Set of images $x_{1}$,..,$x_{n}$ with targets $z_{1}$,...,$z_{n}$}
    \Output{Tree $T$}
    Train the Siamese convolutional network with the sample $(x_i,z_i)$ using the loss \eqref{loss} and $y_{ij}=0 \iff z_i=z_j$.

    Compute the distances $D_{ij}$ between (the embedding of) every pair of images $(x_i,x_j)$ in the Poincar\'e ball.

    $T\longleftarrow$ Neighborjoining($D$)
    \caption{Algorithm for tree reconstruction}
\end{algorithm}

\subsection{Synthetic Data}

We generate synthetic images with a hierarchical tree structure. The algorithm to produce the images is as follows: we split the image in two equal parts. The left half of the image is colored either black or white representing a first branch that splits into a zero or a one. Then, we split the right half of the image into two equal parts, the upper half is colored either white or black. This defines a new branching of the tree into zero or one. One quarter of the image is still left undecided about its color at this stage. In the next step, the left half of the lower right quarter takes either white or black color. Repeating the procedure for five steps, choosing in a spiral way form the sector of the image to color at each step, different classes representing the leaves of a tree are obtained. After five branchings, the remaining area of the image is colored in black and white in equal proportions. The set of all images that can be produced with this algorithm can be thought as leaves of a tree.

\begin{figure}
\centering
\subfloat[\centering 00001]{{\includegraphics[width=.15\textwidth]{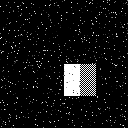} }}%
\subfloat[\centering 00011]{{\includegraphics[width=.15\textwidth]{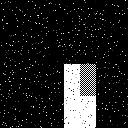} }}%
\subfloat[\centering 01000]{{\includegraphics[width=.15\textwidth]{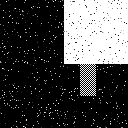} }}%
\subfloat[\centering 01001]{{\includegraphics[width=.15\textwidth]{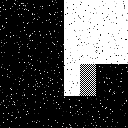} }}%
\subfloat[\centering 11110]{{\includegraphics[width=.15\textwidth]{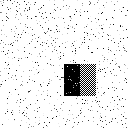} }}%
\subfloat[\centering 10000]{{\includegraphics[width=.15\textwidth]{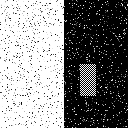} }}%
\caption{Six images with implicit hierarchy.}%
\label{fig: imagenes artificial}%
\end{figure}

\begin{figure}
\centering
\subfloat[\centering Binary tree]{{\includegraphics[scale=0.25]{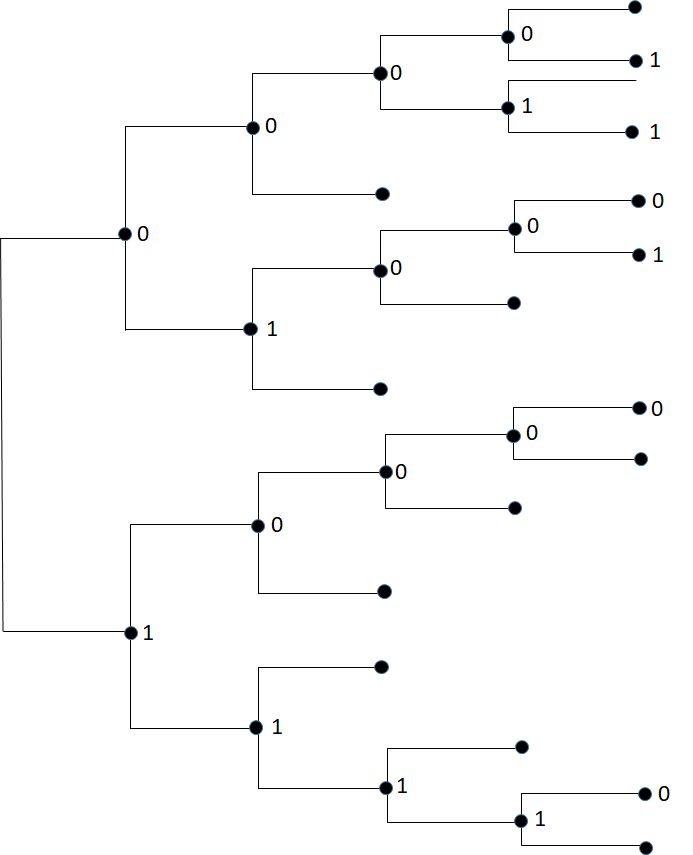} }}%
\qquad
\subfloat[\centering Simplified binary tree]{{\includegraphics[scale=0.25]{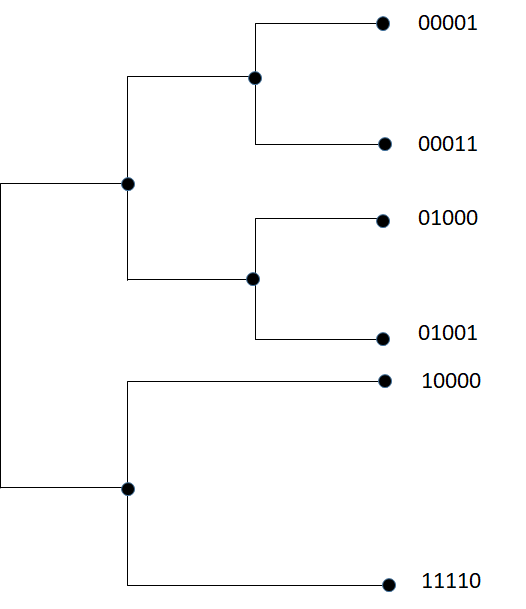} }}%
\caption{Implicit hierarchical structure of the synthetic images.}%
\label{fig: arbol fake}%
\end{figure}

Figure \ref{fig: imagenes artificial} shows six leaves chosen at random. For those leaves, we generate 80 images to train the model and 20 for validation. Noise is added to each of the images by changing the color of 10$\%$ of the pixels to others of different shades of gray. The tree structure representing them is shown in Figure \ref{fig: arbol fake}. The tree (b) is the desired result to be reconstructed.

The model, which aims to find embeddings to represent the data is a Siamese neural network. Each of the twin networks is itself a convolutional net. Using the notation $N @ k \times k$ for a convolutional layer with $N$ filters and a kernel of size $k \times k$, the structure of the convolutional network is as follows:
\begin{itemize}
\item Convolution + ReLU $64 @ 10 \times 10$
\item Max Pool $2 \times 2$
\item Convolution + ReLU $128 @ 7\times7$
\item Max Pool $2\times 2$
\item Convolution + ReLU $256 @ 4 \times 4$
\item Max Pool $2 \times 2$
\item Convolution + ReLU $512 @ 4 \times 4$
\item Max Pool $2 \times 2$
\item Dense Layer 512 + ReLU
\item Dense Layer 2 + Linear
\item $x^* \gets 0.99 \tanh(|x|) \frac{x}{|x|}$
\end{itemize}

Thus, the Siamese neural network is defined using two convolutional networks with ReLU activation function; each of which takes an image and produces a vector within the Poincar\'e disk in two dimensions. The network follows the typical structure of Siamese networks, after each convolutional layer, which doubles the number of filters of the previous one, a pooling layer is placed to reduce the dimension.

To compute the hyperbolic distance between these vectors using the Poincar\'e distance defined in \ref{eqn:poincare distance} we need the output vectors to belong to the (open) ball of radius one; then we apply the following transformation to the vector $x$ obtained as the two-dimensional result of the last dense layer: $x^* = .99 \tanh(|x|) \frac{x}{|x|}$.

\begin{figure}
\centering
\includegraphics[scale=0.25]{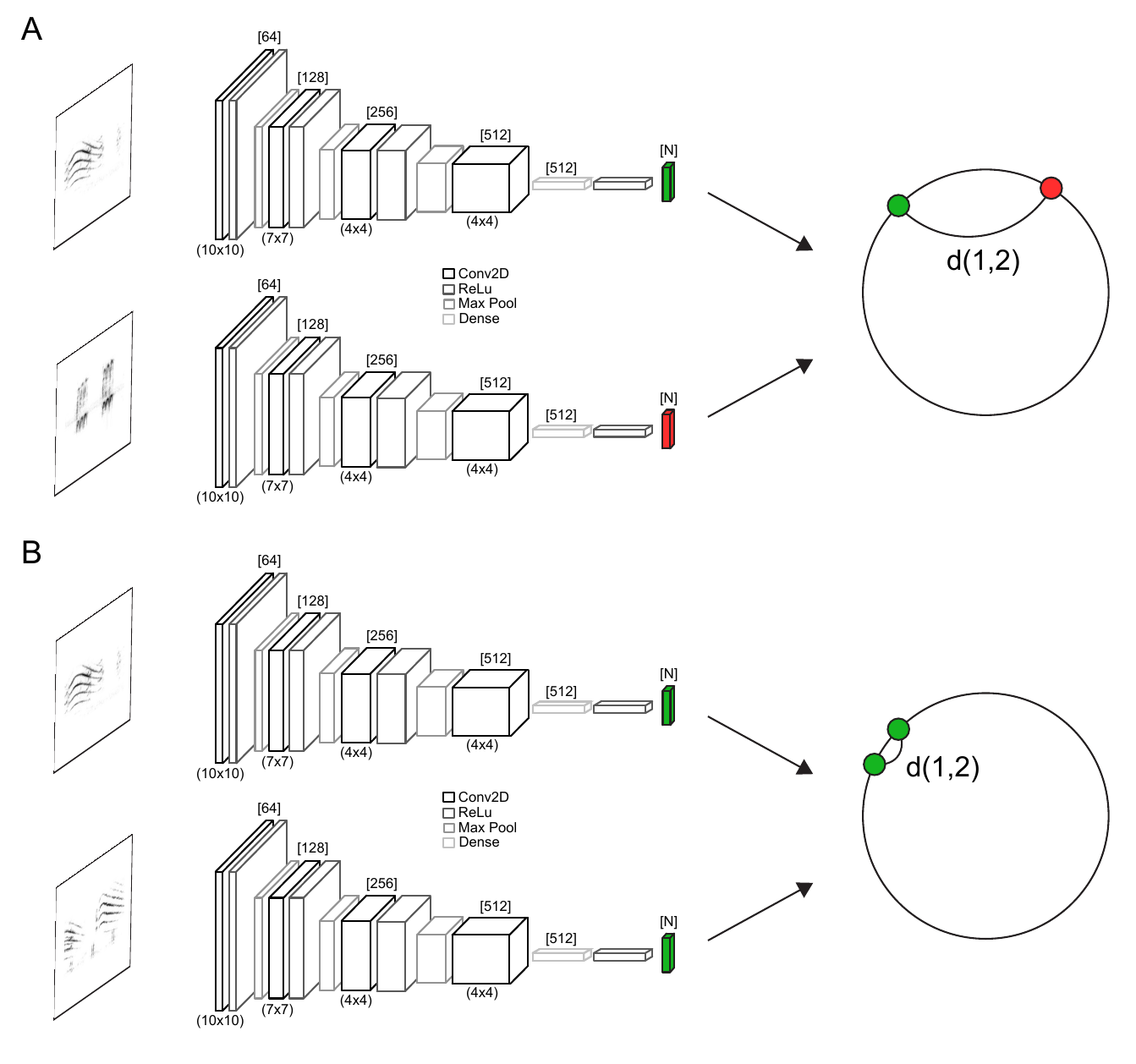}
\caption{Siamese network architecture. The distance between the outputs of each of the networks is computed in the Poincaré ball. The loss function imposes the embedding of different classes to be separated in the Poincaré ball while embeddings of elements of the same class are forced to be close in the Poincaré ball. }
\label{fig: cnn implementation}
\end{figure}

Also, by using the $\tanh$ function, points close to the center remain close, and those with large norm are closer to the boundary of the ball. The new point $x^*$ maintains the direction of $x$. Once both vectors are obtained within the ball of radius one, it is possible to calculate the Poincar\'e distance between them, see Figure \ref{fig: cnn implementation}

Learning is done with a series of mini-batches of 32 pairs of images. The choice of these pairs is not completely random: half of them are positive, meaning that they are chosen so that they belong to the same class, and the other half are negative, meaning that they belonged to different classes. The model is trained for 150 epochs or iterations. The parameter $m$ for the contrastive loss is set to $m=25$. Adam optimizer is used, along with a learning rate parameter of $10^{-6}$.

We compute the embedding ten times. For each of them, the initialization of the network weights is varied following a Gaussian distribution. To measure the performance of the neural network, 10 classes are randomly chosen, allowing repetition. For each class one image is selected, called support image. Then, a test image is chosen, and to compare it with the support images, the convolutional model is applied to each of them. If the Poincaré distance between the test image and the support image of the same class is the lowest then we say that the task is successful. Repeating the process 100 times we obtain a measure of the accuracy of the model as the average of the successful tasks. 

\begin{figure}
\centering
\subfloat{{\includegraphics[width=.3\textwidth]{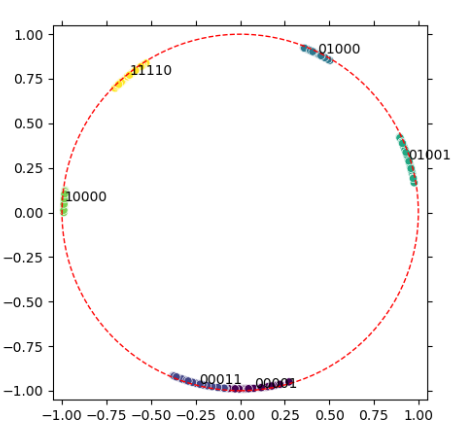} }}
\subfloat{{\includegraphics[width=.3\textwidth]{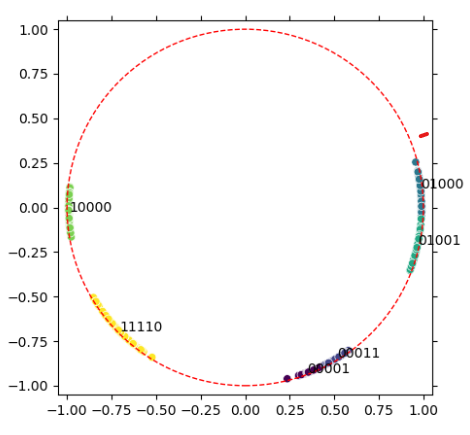} }}
\subfloat{{\includegraphics[width=.3\textwidth]{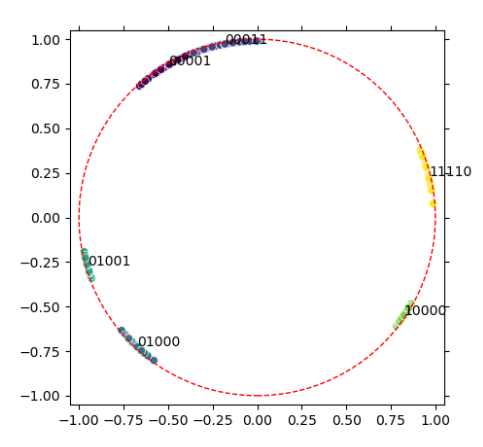} }}

\subfloat{{\includegraphics[width=.3\textwidth]{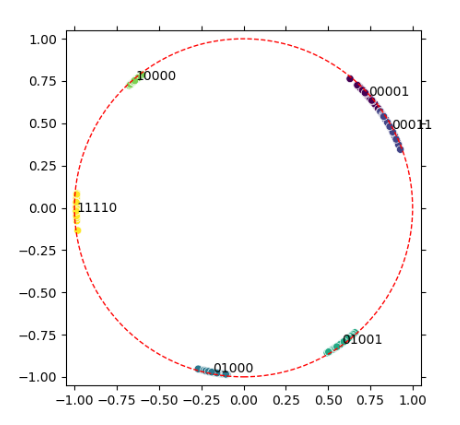} }}
\subfloat{{\includegraphics[width=.3\textwidth]{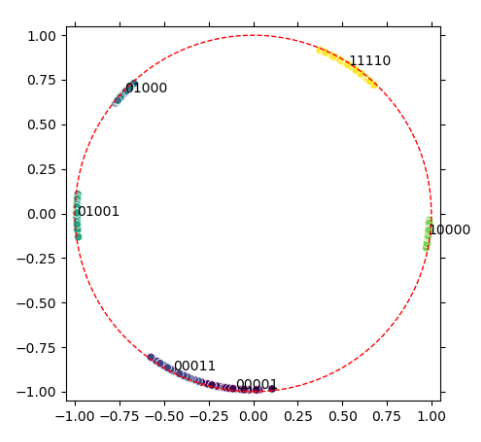} }}
\subfloat{{\includegraphics[width=.3\textwidth]{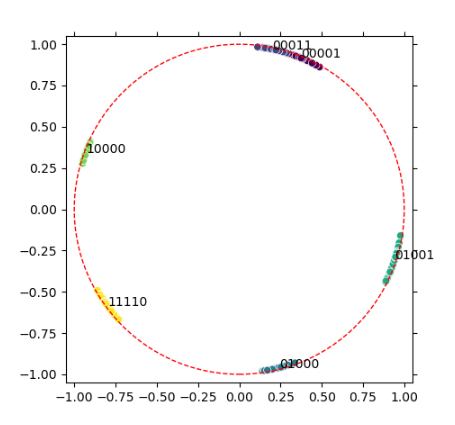} }}
\caption{Embeddings in the Poincar\'e disk. Up: runs 1,2 and 7. Down: runs 8,9 and 10.}
\label{fig: embeddings fake2}
\end{figure}

The embeddings obtained for this data can be seen in Figure \ref{fig: embeddings fake2}. The neural network successfully separates the different classes and achieved $98.2\%$ accuracy in training and $97\%$ in validation. The results for each run are shown in Table \ref{table: accuracy fake}.

\begin{center}
\begin{table}
\begin{center}
\begin{tabular}{|lll|}
\hline
\multicolumn{3}{|c|}{Accuracy}
\\ \hline
\multicolumn{1}{|l|}{} & \multicolumn{1}{l|}{Training} & Validation \\ \hline
\multicolumn{1}{|l|}{1} & \multicolumn{1}{l|}{0.99} & 0.98 \\ \hline
\multicolumn{1}{|l|}{2} & \multicolumn{1}{l|}{0.99} & 0.97 \\ \hline
\multicolumn{1}{|l|}{3} & \multicolumn{1}{l|}{0.98} & 0.97 \\ \hline
\multicolumn{1}{|l|}{4} & \multicolumn{1}{l|}{0.99} & 0.99 \\ \hline
\multicolumn{1}{|l|}{5} & \multicolumn{1}{l|}{0.97} & 0.95 \\ \hline
\multicolumn{1}{|l|}{6} & \multicolumn{1}{l|}{0.96} & 0.97 \\ \hline
\multicolumn{1}{|l|}{7} & \multicolumn{1}{l|}{0.99} & 0.96 \\ \hline
\multicolumn{1}{|l|}{8} & \multicolumn{1}{l|}{0.98} & 0.95 \\ \hline
\multicolumn{1}{|l|}{9} & \multicolumn{1}{l|}{0.98} & 0.99 \\ \hline
\multicolumn{1}{|l|}{10} & \multicolumn{1}{l|}{0.99} & 0.97 \\ \hline
\end{tabular}
\end{center}
\caption{Accuracy of the model applied to synthetic data.}
\label{table: accuracy fake}
\end{table}
\end{center}

\begin{figure}
\centering
\subfloat{{\includegraphics[angle=-90, scale=0.25]{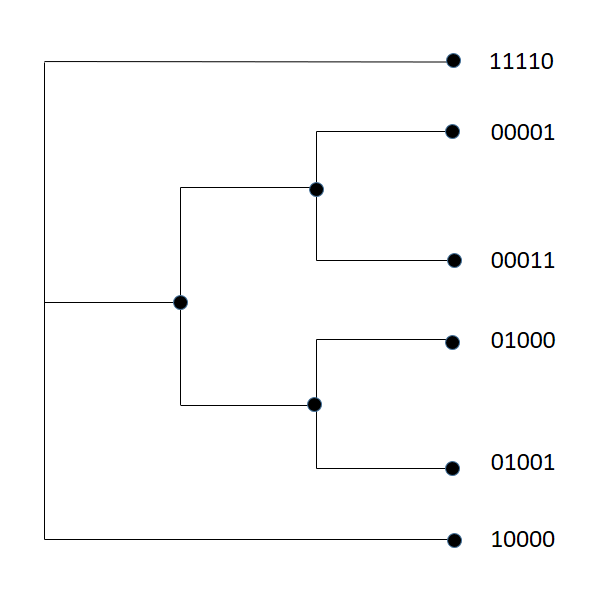} }}
\qquad
\subfloat{{\includegraphics[angle=-90, scale=0.25]{imagenes/arbolfakesimple2.png} }}
\caption{Reconstructed trees with NJ. Left: tree without root obtained in $90\%$ of the runs; Right: rooted tree ({midpoint-rooting}) obtained in $70\%$ of the runs.}
\label{fig: arbol fake reconstruido}
\end{figure}

From the representation of the data in this space, the Chebyshev center of each class is computed using the Poincar\'e distances. Thus, each class has a representative in the embedding space. Using these representatives, the Poincar\'e distance matrix between the classes is computed. Then, we apply NJ algorithm to reconstruct the trees. The result is that in $90\%$ of the runs, tree (a) in Figure \ref{fig: arbol fake reconstruido} is obtained, therefore, each pair of images is correctly identified. The algorithm is robust for this dataset. After applying midpoint rooting to choose the root, in $70\%$ of the runs tree (b) is obtained. Therefore, $70\%$ of the times, the entire hierarchical structure of Figure \ref{fig: arbol fake} is recovered.

\subsection{Sonograms}

We work with sonograms of syllables from the songs of each of the following species: \textit{Stizoptera bichenovii, Poephila acuticauda, Lagonosticta senegala, Lonchura striata, Amandava subflava, Uraeginthus bengalus}. The songs from which the representative syllables are extracted have been taken from the \textit{Xeno-canto} database \cite{xeno}. We selected these species because a recent study compared their phylogenetic organization with the one emerging from comparing the songs by pre-selecting certain acoustic properties \cite{rivera_woolley2023}. The method we present here does not require a pre-specification of properties to carry out the reconstruction.

The complete set of audio files is first converted to mono and down-sampled to 22050 Hz. A 12th-order Butterworth FIR-type band-pass filter is applied with cut-off frequencies between 1 and 8 kHz. Noise reduction is then performed using spectral gating, a technique that applies a noise gate by setting a threshold for each frequency band, suppressing noise components below that threshold. To implement this, we estimate stationary noise thresholds using a signal and noise threshold set at 1.5 standard deviations above the mean, followed by noise suppression at a reduction ratio of 95\%.

Next, the audio files are segmented into 1-second chunks with a 0.5-second overlap, padding with zeros when the chunk is shorter than 1 second, but discarding chunks that required more than 0.25 seconds of padding. Spectrograms for each chunk are generated using the Short-Time Fourier Transform (STFT) with a Hamming window. The window size and hop-size are adapted to obtain spectrograms with a shape of (128, 256). A clipping operation is applied to limit the values to a range between 1.5 times the median (values below this are set to zero) and the maximum. The logarithm of the values is then computed for each element in the spectrogram matrix. Finally, the spectrograms are rescaled to values between 0 and 1. The resulting spectrograms are exported as images at 50 dpi resolution for use as inputs to the Siamese networks.

Although the files from Xeno-Canto are labeled with the presence of a particular species, the audio recordings containe a variety of noise and vocalizations, which may belong to the labeled species or to others present in the recordings. Consequently, the data is considered weakly labeled. The challenge is to extract the specific vocalizations corresponding to the labeled species. To address this, we manually select audio segments that contained vocalizations of the species of interest. We cross-referenced these selections with audiovisual material from YouTube to increase confidence in the species-specific vocalizations and identified recurring patterns in the Xeno-Canto files. We then select audio segments corresponding to one unique syllable type for each species, both from the Xeno-Canto database and the YouTube recordings, yielding a total of 568 segments (poephila + acuticauda: 51; stizoptera + bichenovii: 133; lonchura + striata: 159; lagonosticta + senegala: 69; uraeginthus + bengalus: 69; amandava + subflava: 87). These selected segments are used for training and evaluating the network, chosen in advance to avoid data leakage.

\begin{figure}
\centering
\subfloat[\centering Species Stizoptera bichenovii]{{\includegraphics[width=.3\textwidth]{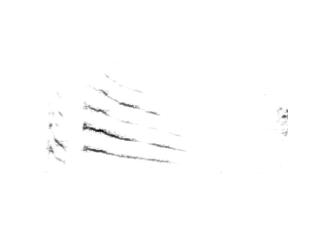} }}
\subfloat[\centering Species Poephila acuticauda]{{\includegraphics[width=.3\textwidth]{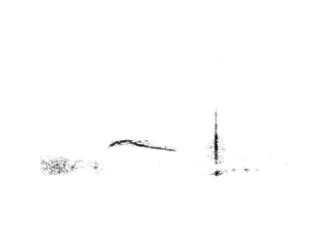} }}%
\subfloat[\centering Species Lagonosticta senegala]{{\includegraphics[width=.3\textwidth]{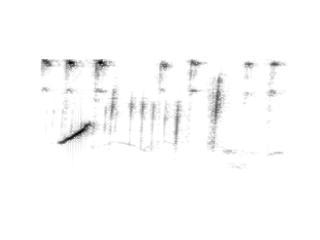} }}%

\subfloat[\centering Species Lonchura striata]{{\includegraphics[width=.3\textwidth]{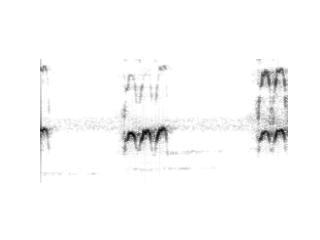} }}%
\subfloat[\centering Species Uraeginthus bengalus]{{\includegraphics[width=.3\textwidth]{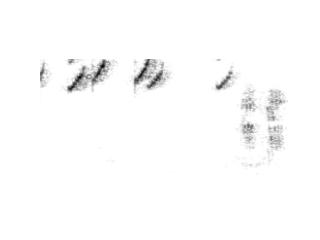} }}%
\subfloat[\centering Species Amandava subflava]{{\includegraphics[width=.3\textwidth]{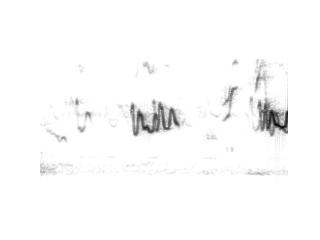} }}%
\caption{Six sonograms of representatives of different bird species.}%
\label{fig: sonograms}%
\end{figure}

Sonograms of the syllables are obtained for each of these species. An example from each class is shown in Figure \ref{fig: sonograms}. The dataset consists of 568 images, 80$\%$ of them are used to train the network and the remaining 20$\%$ for validation. Additionally, the data is augmented by applying transformations: rotations, translations, and zoom.

The structure of the convolutional network used to obtain the embeddings is similar to the one used with the synthetic data. The only difference is that the final output layer has three dimensions instead of two.
\begin{itemize}
\item Convolution + Relu $64 @ 10 \times 10$
\item Max Pool $2 \times 2$
\item Convolution + Relu $128 @ 7\times7$
\item Max Pool $2\times 2$
\item Convolution + Relu $256 @ 4 \times 4$
\item Max Pool $2 \times 2$
\item Convolution + Relu $512 @ 4 \times 4$
\item Max Pool $2 \times 2$
\item Dense Layer 512 + Relu
\item Dense Layer 3 + Linear
\item $x^* \gets 0.99 \tanh(|x|) \frac{x}{|x|}$
\end{itemize}

The Siamese neural network uses two convolutional networks, each of which takes a sonogram and produces a point $x^*$ within the Poincar\'e ball in dimension three.

Similarly to the synthetic data, learning is performed with a series of mini-batches of 32 pairs of images where half of them correspond to sonograms of the same class and the other half to sonograms of different classes. The model require 2500 epochs to converge.

The average result obtained is $88.3\%$ accuracy for the training set and $86.3\%$ for the validation set. In Table \ref{table: accuracy sonograms}, the accuracy values for each of the ten runs is shown.

\begin{table}[]
\begin{center}
\begin{tabular}{|lll|}
\hline
\multicolumn{3}{|c|}{Accuracy} \\ \hline
\multicolumn{1}{|l|}{} & \multicolumn{1}{l|}{Training} & Validation \\ \hline
\multicolumn{1}{|l|}{1} & \multicolumn{1}{l|}{0.91} & 0.89 \\ \hline
\multicolumn{1}{|l|}{2} & \multicolumn{1}{l|}{0.93} & 0.91 \\ \hline
\multicolumn{1}{|l|}{3} & \multicolumn{1}{l|}{0.85} & 0.82 \\ \hline
\multicolumn{1}{|l|}{4} & \multicolumn{1}{l|}{0.89} & 0.87 \\ \hline
\multicolumn{1}{|l|}{5} & \multicolumn{1}{l|}{0.88} & 0.86 \\ \hline
\multicolumn{1}{|l|}{6} & \multicolumn{1}{l|}{0.89} & 0.86 \\ \hline
\multicolumn{1}{|l|}{7} & \multicolumn{1}{l|}{0.79} & 0.81 \\ \hline
\multicolumn{1}{|l|}{8} & \multicolumn{1}{l|}{0.9} & 0.87 \\ \hline
\multicolumn{1}{|l|}{9} & \multicolumn{1}{l|}{0.91} & 0.89 \\ \hline
\multicolumn{1}{|l|}{10} & \multicolumn{1}{l|}{0.88} & 0.85 \\ \hline
\end{tabular}
\end{center}
\caption{Accuracy for the model applied to sonogram data.}
\label{table: accuracy sonograms}
\end{table}

\begin{figure}
    \centering
    {{\includegraphics[width=.3\textwidth]{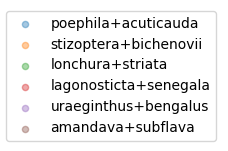} }}%
    {{\includegraphics[width=.6\textwidth]{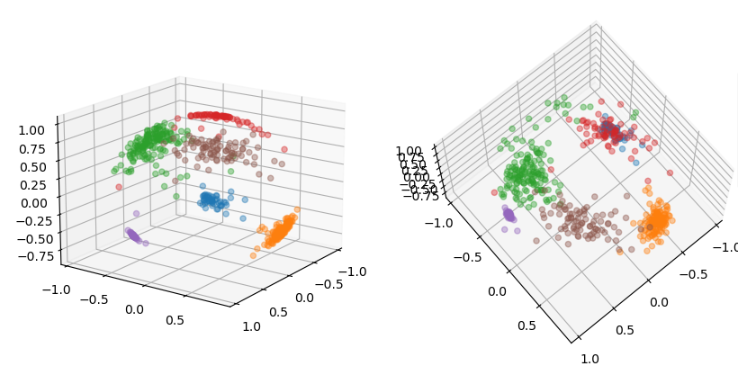} }}%
    \caption{Sonograms embedding.}
    \label{fig: embedding son 1}
\end{figure}

The embeddings obtained in the Poincar\'e space in the first run can be observed in Figure \ref{fig: embedding son 1}.

Based on the data representation in this space, we compute the Chebyshev center of each class with respect the Poincar\'e distance. Using this matrix of Poincar\'e distances between the centers, the NJ algorithm is applied.

\begin{figure}
    \centering
    \subfloat[\centering]
    {{\includegraphics[angle=-90, width=.25\textwidth]{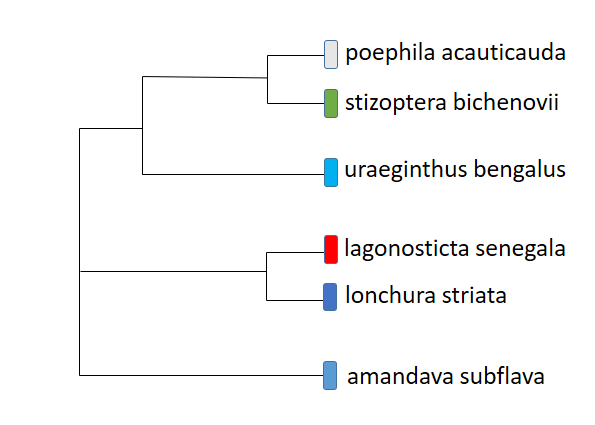} }}
    \subfloat[\centering]
    {{\includegraphics[angle=-90, width=.25\textwidth]{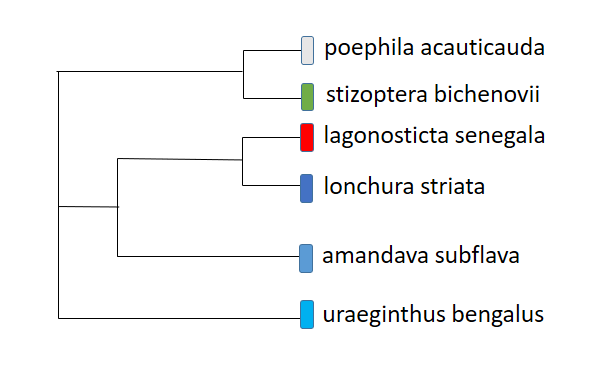} }}
    \subfloat[\centering]
    {{\includegraphics[angle=-90, width=.25\textwidth]{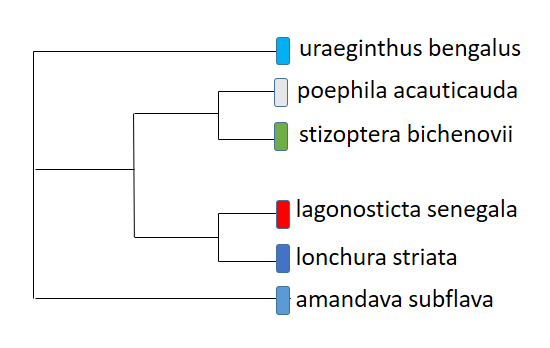} }}%
    \caption{Trees generated by NJ: (a) Unrooted tree generated from sonograms for 70\% of the runs; (b) Unrooted tree generated from sonograms for 20\% of the runs.; (c)  Unrooted tree generated from sonograms for 10\% of the runs.}
    \label{fig: arbol fil sin raiz}
\end{figure}

\begin{figure}
\centering
    \subfloat[\centering]
    {{\includegraphics[angle=-90,width=.22\textwidth]{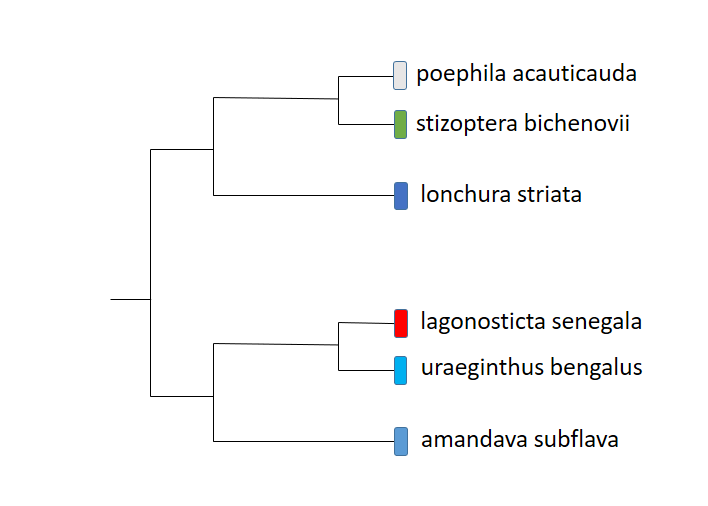}}}
    \subfloat[\centering]{{\includegraphics[angle=-90, width=.22\textwidth]{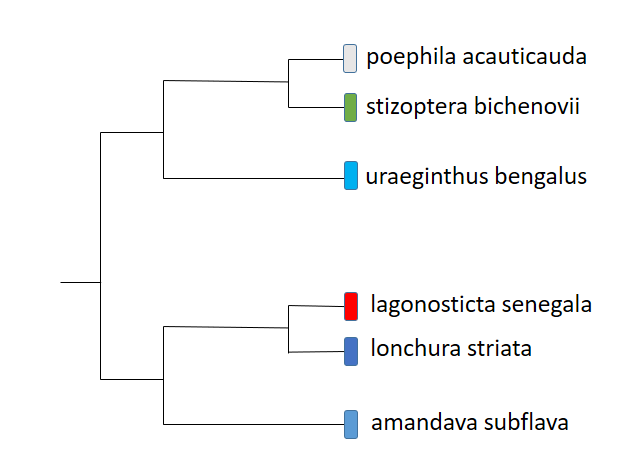} }}
    \subfloat[\centering]{{\includegraphics[angle=-90, width=.22\textwidth]{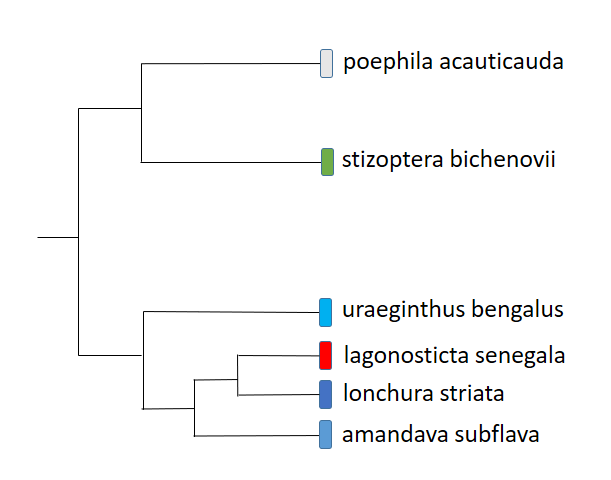} }}
    \subfloat[\centering]{{\includegraphics[angle=-90, width=.22\textwidth]{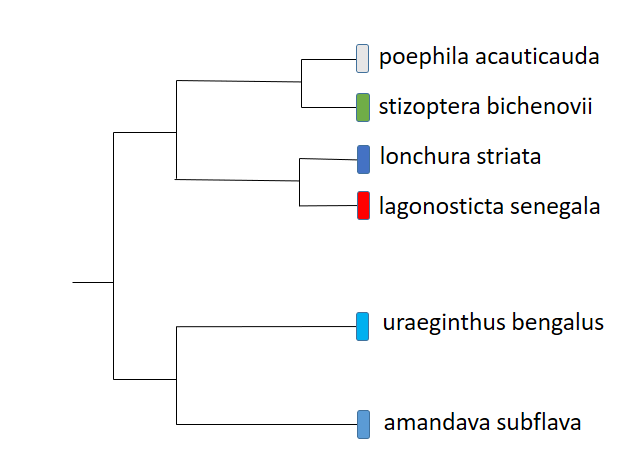} }}
    \caption{$(a)$ Phylogenetic tree of bird species. $(b)-(d)$ Trees with roots using midpoint rooting and NJ. Tree (b) was obtained in 60\% of the runs. Tree (c) was obtained in 30\% of the runs. Tree (d) was obtained in 10\% of the runs.}
    \label{fig: arboles con raiz}
\end{figure}

The unrooted trees obtained from the ten runs can be seen in Figure \ref{fig: arbol fil sin raiz}. Subsequently, each of the trees is rooted using midpoint rooting. The results can be found in Figure \ref{fig: arboles con raiz}. In $60\%$ of the runs the same tree $(b)$ is obtained. Its structure presents similarities with the phylogenetic tree. First, \textit{Poephila acuticauda} and \textit{Stizoptera bichenovii} are neighbors, as in the phylogenetic tree. \textit{Amandava subflava} and \textit{Lagonosticta senegala} derive from the same common ancestor for both trees. The main difference between these two trees lies in the placement of \textit{Lonchura striata} and \textit{Uraeginthus bengalus}. The former shares a parent with \textit{Lagonosticta senegala} in the song tree, while in the phylogenetic tree, the closest common ancestor is the parent of \textit{Poephila acuticauda} and \textit{Stizoptera bichenovii}. In contrast, for the song tree, \textit{Uraeginthus bengalus} occupies the place that \textit{Lonchura striata} has in the phylogenetic tree.

It is noteworthy that the phylogenetic tree differs from the tree generated from the behavior in the position of a species: \textit{Uraenginthus bengalus}. This species has some characteristics that warrant a review of the data used. In particular, the female sings, although with a less complex song than the male. There is a possibility that our dataset included examples of female songs for this species, which would compromise the classification. It would be interesting to asses this hypothesis.

The remaining two song trees have structures with more differences compared to the phylogenetic one. In all runs, \textit{Poephila acuticauda} and \textit{Stizoptera bichenovii} are neighbors, and \textit{Lagonosticta senegala} and \textit{Lonchura striata} are also neighbors.


\section{Conclusion}
\label{conclusion}

In recent years, we have witnessed the impressive growth of a new paradigm in science. One that, in some way, contrasts with the ``interpretative'' attitude toward phenomena that has been characteristic of natural sciences in the last centuries. This new approach, based on data, establishes complex correlations between data without aiming to elucidate mechanisms or employ easily interpretable variables. In the context of this work, we explore what kind of phylogenetic inferences can be drawn from data without the need for prior interpretative processing of the characteristics from which the task is performed. 
We discuss two examples of how to unveil phylogenia from data conveyed as images. One where the ``evolutionary'' process is simulated computationally, and another one where the data comes from the natural world (songs produced by individuals of closely related species). We take the second example because neural networks were recently employed to make phylogenetic inferences from birdsong. Opposite to our case, in that case the variables used are ``interpretable''. In fact, they are intuitive acoustic features easily recognizable in the song spectrograms. This leads us to the second example.

The first example is synthetic and is used to test the method since in this case, the ``phylogeny'' is known in advance. We observe that the algorithm is capable to find the hierarchical structure in the data besides having only information about the leafs of the tree. That is, the algorithm reconstructs the sub-tree determined by the leafs for which we have examples in the dataset. The only input required by the algorithm is images that encode, in some way, relevant characteristics to determine the hierarchy. We remark that we don't need to known what are those characteristics. They just need to be present in the data. In view of this, we can expect the algorithm to be able to reconstruct, in real situations, phylogenetic trees departing from images that contain relevant phylogenetic characteristics. Moreover, this leads as to the natural problem of inferring if a set of characteristics contain phylogenetic information or not and even more, if there is an additional hierarchical structure besides the phylogeny or if it is the case that it is just that.

The second example is very subtle. Songs, in the species that are studied in this work, are learned. Therefore, some features are conditioned genetically, while others are learned from other individuals. Then, choosing the right interpretable variables is challenging: we have to choose those that are more likely to be genetically programmed. A data driven approach might identify the similarities without pre-determination of the features.

In this work, we train a Siamese neural network with a metric which is designed to unveil hierarchies. The result is that statistically, it is capable to infer the tree structure behind the species which is pretty similar to the phylogenetic one. 

In any data driven study, data preparation is a key aspect of the process and might induce specific biases. In our application, the selection of audio segments was performed manually, using a single syllable type as the representative song pattern for each species. Yet, the Siamese network demonstrated strong performance in species recognition, yielding non-trivial results. The generated trees exhibited consistent robustness across multiple runs, further underscoring the method's effectiveness. Future enhancements, such as incorporating laboratory-controlled data or automated methods for labeling audio segments, could expand the applicability of this approach. Nevertheless, the current results highlight both the potential and robustness of the method, as it was evaluated on data that was selected prior to model training, ensuring it remained agnostic and unaffected by data leakage.


\bibliographystyle{plain}
\bibliography{bibliography}

\end{document}